\documentstyle[twocolumn,floats,aps]{revtex}

\begin{document}
\draft

\twocolumn[\hsize\textwidth%
\columnwidth\hsize\csname@twocolumnfalse\endcsname

\title{\bf A Discrete Solid-on-Solid Model for Nonequilibrium Growth
Under Surface Diffusion Bias}

\author{S. Das Sarma and P. Punyindu}
\address{Department of Physics, University of Maryland, College Park, 
MD 20742-4111}

\date{\today}
\maketitle

\begin{abstract}
A limited mobility nonequilibrium solid-on-solid dynamical model 
for kinetic surface growth is introduced as a simple description for the
morphological evolution of a growing interface under random vapor
deposition and surface diffusion bias conditions. Large scale stochastic
Monte Carlo simulations using a local coordination dependent
instantaneous relaxation of the deposited atoms produce complex surface
morphologies whose dynamical evolution is not consistent with any of the
existing continuum dynamical surface growth equations. 
Critical exponents for coarsening and
roughening dynamics of the morphological evolution 
are quantitatively calculated using large scale simulations.
\end{abstract}

\vskip 1pc]
\narrowtext

An atom moving on a free surface is known to encounter an additional 
potential barrier, often called a surface diffusion bias \cite{1} 
(or an Ehrlich \cite{2} - Schwoebel \cite{3} barrier), as it approaches
a step from the upper terrace -- there is no such extra barrier for an
atom approaching the step from the lower terrace (the surface step
separates the upper and the lower terraces). Since this diffusion bias
makes it preferentially more likely for an atom to attach itself to the
upper terrace than the lower one, it leads to mound (or pyramid) - type
structures on the surface under growth conditions as deposited atoms
are probabilistically less able to come down from upper to lower
terraces. This dynamical growth behavior under a surface diffusion bias
is sometimes called an ``instability'' because a flat two dimensional 
surface growing under a strong surface diffusion bias is unstable
toward three dimensional mound/pyramid formation. 
There has been a great deal of recent interest 
\cite{1,4,5,6,7,8,9,10,11,12,13,14,15,16,17,18,19,20} in the 
morphological evolution of growing interfaces under nonequilibrium 
growth conditions in the presence of such surface diffusion bias. 
In this Letter we introduce, what we believe to be, the {\it minimal}
nonequilibrium atomistic growth model for 
ideal molecular beam epitaxial -
type random vapor deposition growth under a surface diffusion
bias. Extensive stochastic simulation results presented in this paper
establish the morphological evolution of a surface growing under 
diffusion bias conditions to be rather complex. 
\begin{figure}

 \vbox to 3.cm {\vss\hbox to 6cm
 {\hss\
   {\includegraphics{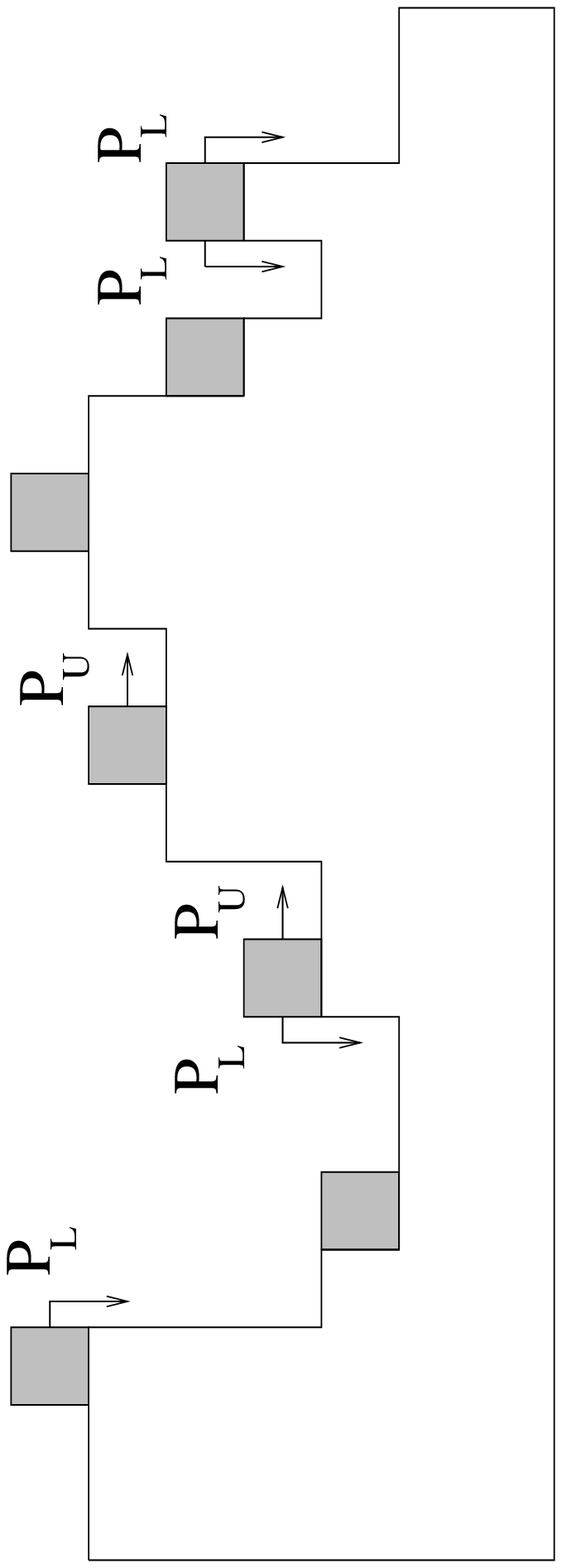}
   }
  \hss}
 }

 \vbox to 4.5cm {\vss\hbox to 6cm
 {\hss\
   {\includegraphics{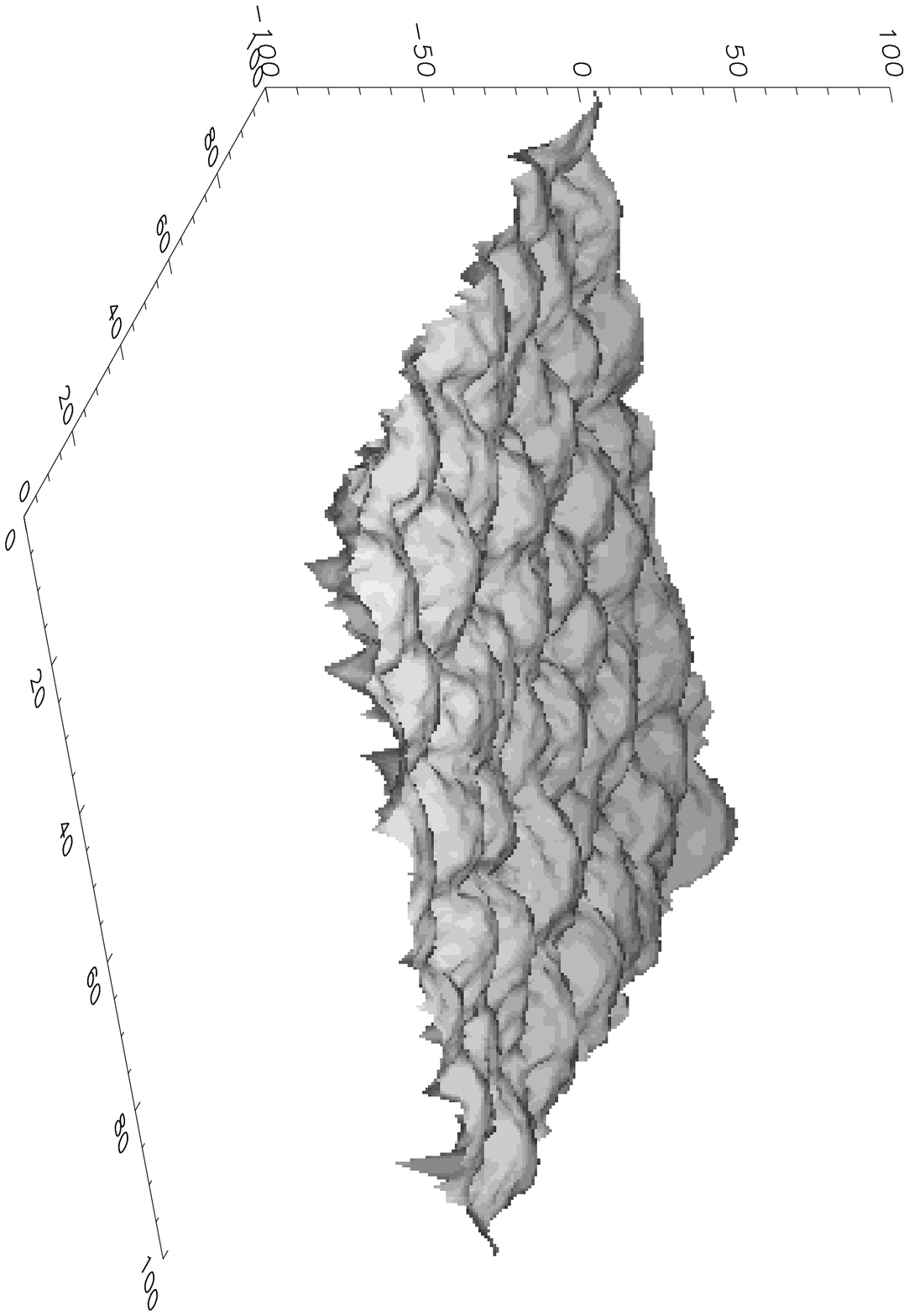}
   }
  \hss}
 }

 \vbox to 4.5cm {\vss\hbox to 6cm
 {\hss\
   {\includegraphics{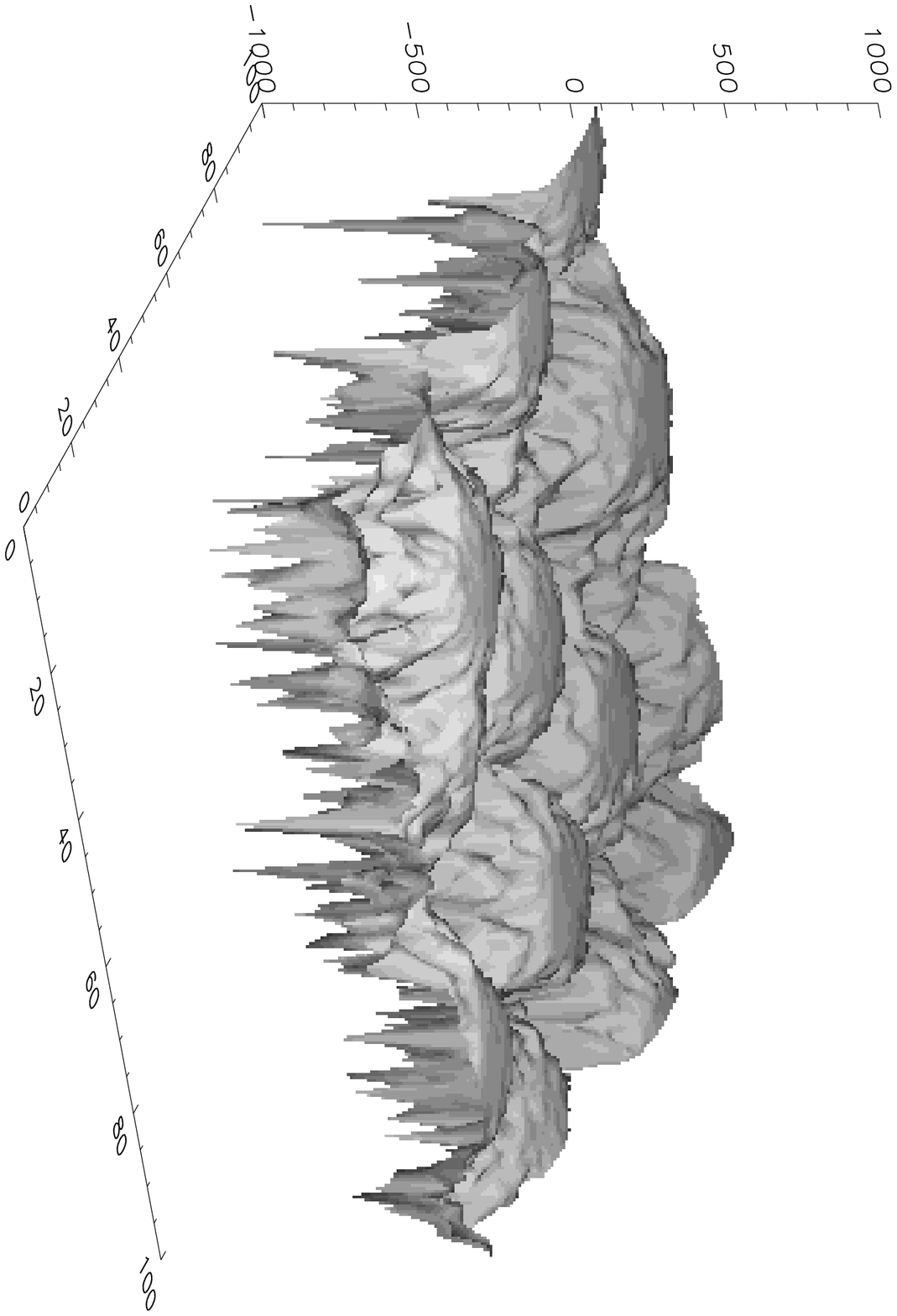}
   }
  \hss}
 }

\caption{
(a) Schematic configuration defining growth rules
in 1+1 dimensions.
(b) Morphologies ($P_U=1; P_L=0.5$) for $L=100 \times 100$ at
$10^3$ ML and (c) $10^6$ ML.
}
\end{figure}
Various critical growth exponents \cite{21,22}, which
asymptotically describe the large-scale dynamical evolution of the
kinetically growing surface in our minimal discrete growth model, are
inconsistent with {\it all} the proposed continuum theories for
nonequilibrium surface growth under diffusion bias conditions. Our
results lead to the conclusion that a continuum description for
nonequilibrium growth under a surface diffusion bias does not
exist and may require a theoretical formulation which is substantially
different from the ones currently existing in the literature.

In Fig. 1(a) we schematically describe our solid-on-solid (SOS) 
nonequilibrium growth model : (1) Atoms are deposited randomly (with
an average rate of 1 layer/unit time, which defines the unit of time in
the growth problem -- the length unit is the lattice spacing taken to be
the same along the substrate plane and the growth direction) and
sequentially on the surface starting with a flat substrate; (2) a
deposited atom is incorporated instantaneously if it has at least one
lateral nearest - neighbor atom (i.e. if it has a coordination of 2 or
more since there must always be an atom underneath to satisfy the SOS
constraint); (3) singly coordinated deposited atoms (i.e. the ones
without any lateral neighbors) could instantaneously relax to a
neighboring site within a diffusion length of $l$ provided the
neighboring site of incorporation has a higher coordination than the
original deposition site; (4) the instantaneous relaxation process is
constrained by two probabilities $P_L$ and $P_U$ (with $0 \leq P_L,P_U
\leq 1$) where $P_{L(U)}$ is the probability for the atom to attach itself
to the lower(upper) terrace after relaxation (note that a ``terrace'' here
could be just one other atom). The surface diffusion bias is implemented
in our model by simply taking $P_U > P_L$, making 
it more likely for atoms to attach to the upper terrace.
Under the surface diffusion bias, therefore, 
an atom deposited at the top of a
step edge feels a barrier (whose strength is controlled by $P_U-P_L$) in
coming down compared with an atom at the lower terrace attaching itself
to the step. Our surface diffusion bias model is 
well-defined for any value of the
diffusion length $l$ including the most commonly studied situation of
nearest - neighbor relaxation ($l=1$).   
We have carried out extensive simulations
both in $1+1$ and $2+1$ dimensions (d) varying $P_L$, $P_U$ as well as
$l$, also including in our simulations the inverse situation (the
so-called `negative' bias condition) with $P_L > P_U$ so that deposited
atoms preferentially come down attaching themselves to lower steps
producing in the process a smooth growth morphology.
Because of lack of space we do not present our negative bias results
here except to note that the smooth dynamical growth morphology
under our negative bias model obeys the generic Edwards - Wilkinson
universality \cite{1,21,22}.

Before presenting our numerical results we point out two important
features of our growth model : (1) For $P_L = P_U = 1$ our model reduces
to the one introduced in ref. 23 (and studied extensively 
\cite{21,22,23,24,25,26,27}
in the literature) as a minimal model for molecular beam epitaxy in the
absence of any diffusion bias; (2) we find, in complete
agreement with earlier findings \cite{23,24} in the absence of diffusion
bias, that the diffusion length $l$ is an {\it irrelevant} variable
which does not affect any of our calculated critical growth exponents
(but does affect finite size corrections -- increasing $l$ requires a
concomitant increase in the system size to reduce finite size effects).
We, therefore, mostly present our $l=1$ simulation results here 
emphasizing that our critical exponents are independent of $l$ provided
finite size effects are appropriately accounted for. Our calculated 
exponents are also independent of the precise values of $P_U$ and
$P_L$ as found in ref. 23 and 24 for the $P_U=P_L$ case.

In Fig.1 we show representative d=2+1 simulated growth
morphologies for our positive bias model. The diffusion 
bias produces mounded structures which are visually statistically scale
invariant only on length scales much larger 
(or much smaller) than the typical mound size.
\begin{figure}

 \vbox to 4.1cm {\vss\hbox to 6cm
 {\hss\
   {\includegraphics{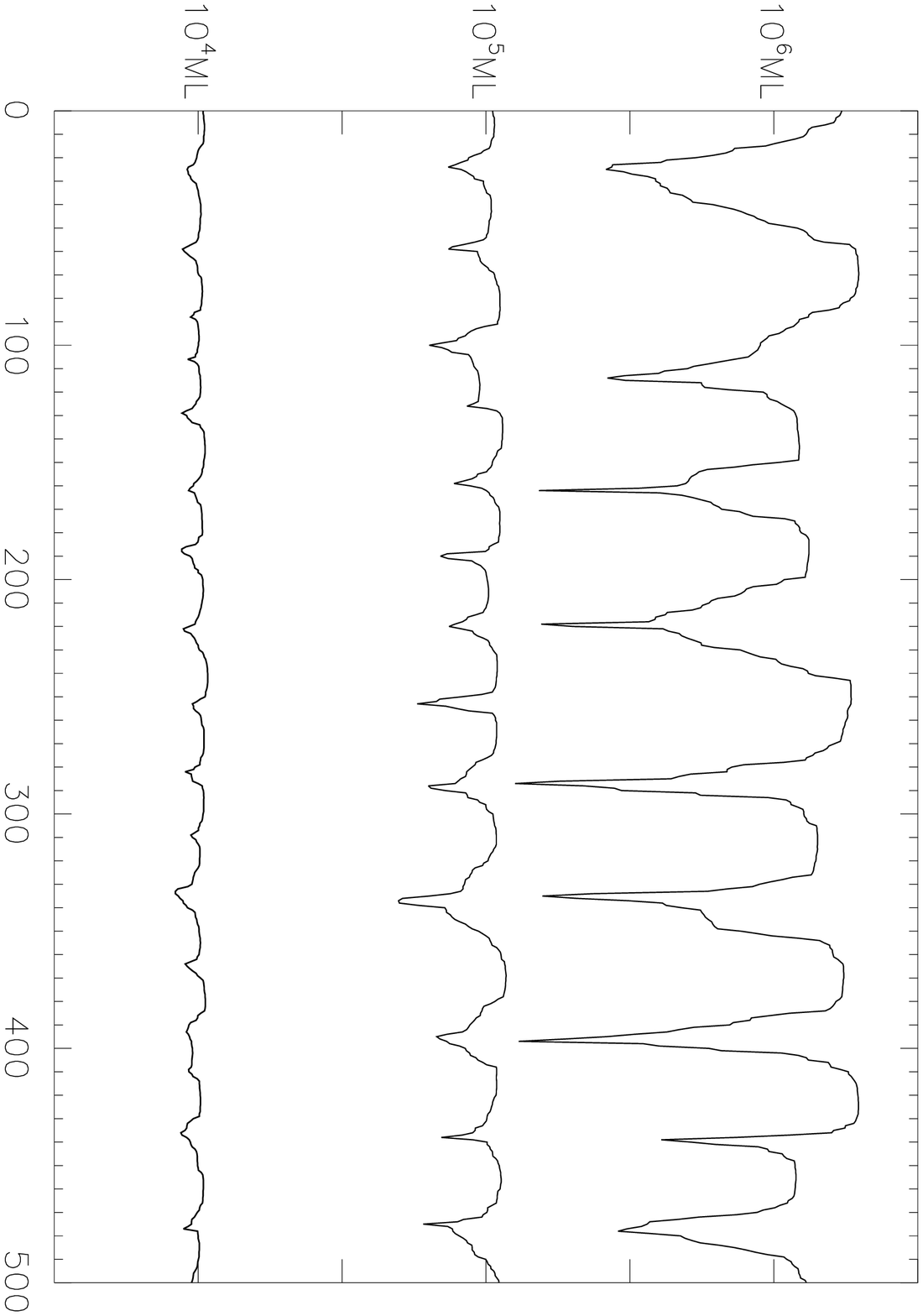}
   }
  \hss}
 }

 \vbox to 4.8cm {\vss\hbox to 6cm
 {\hss\
   {\includegraphics{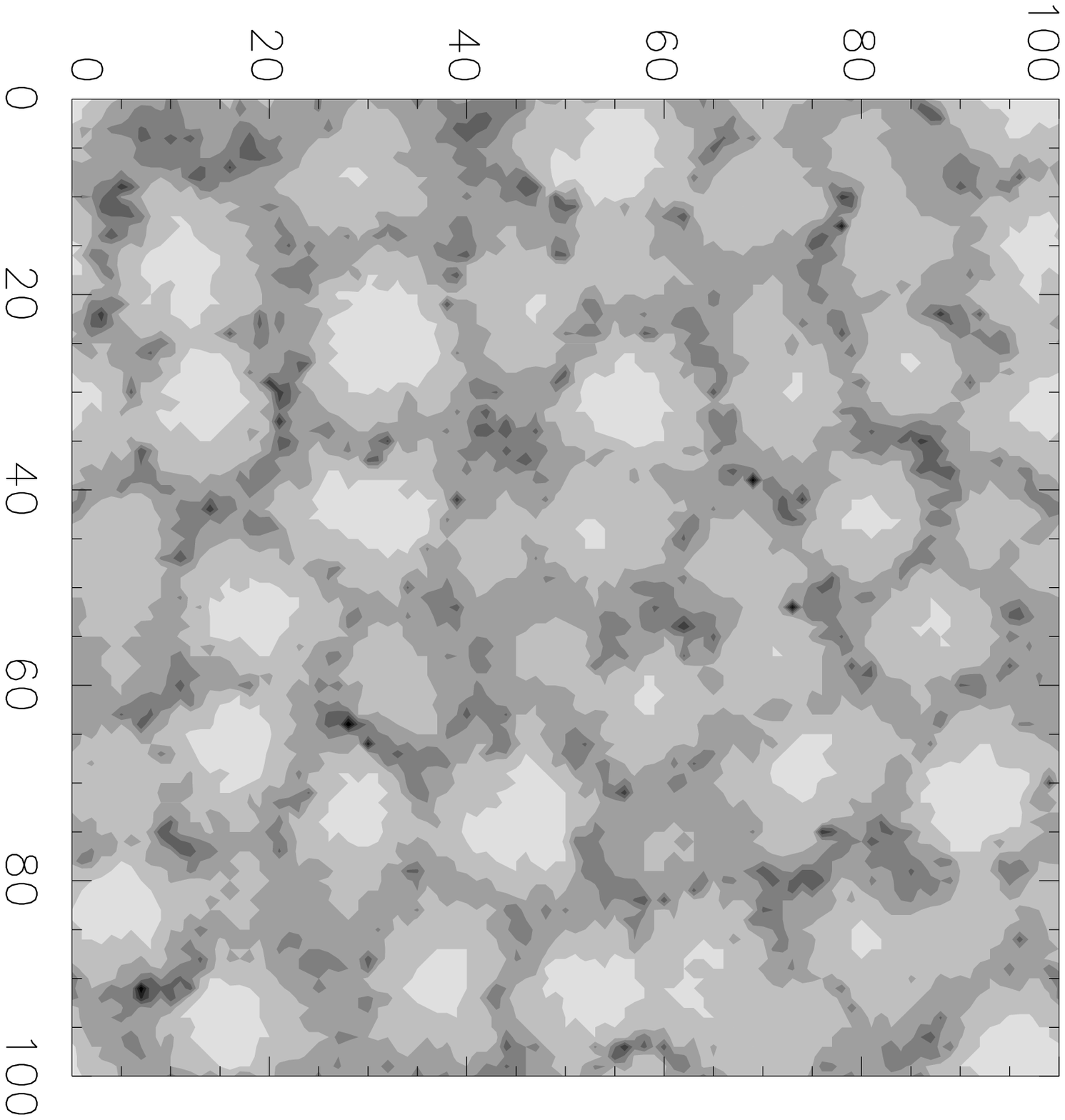}
   }
  \hss}
 }

 \vbox to 4.8cm {\vss\hbox to 6cm
 {\hss\
   {\includegraphics{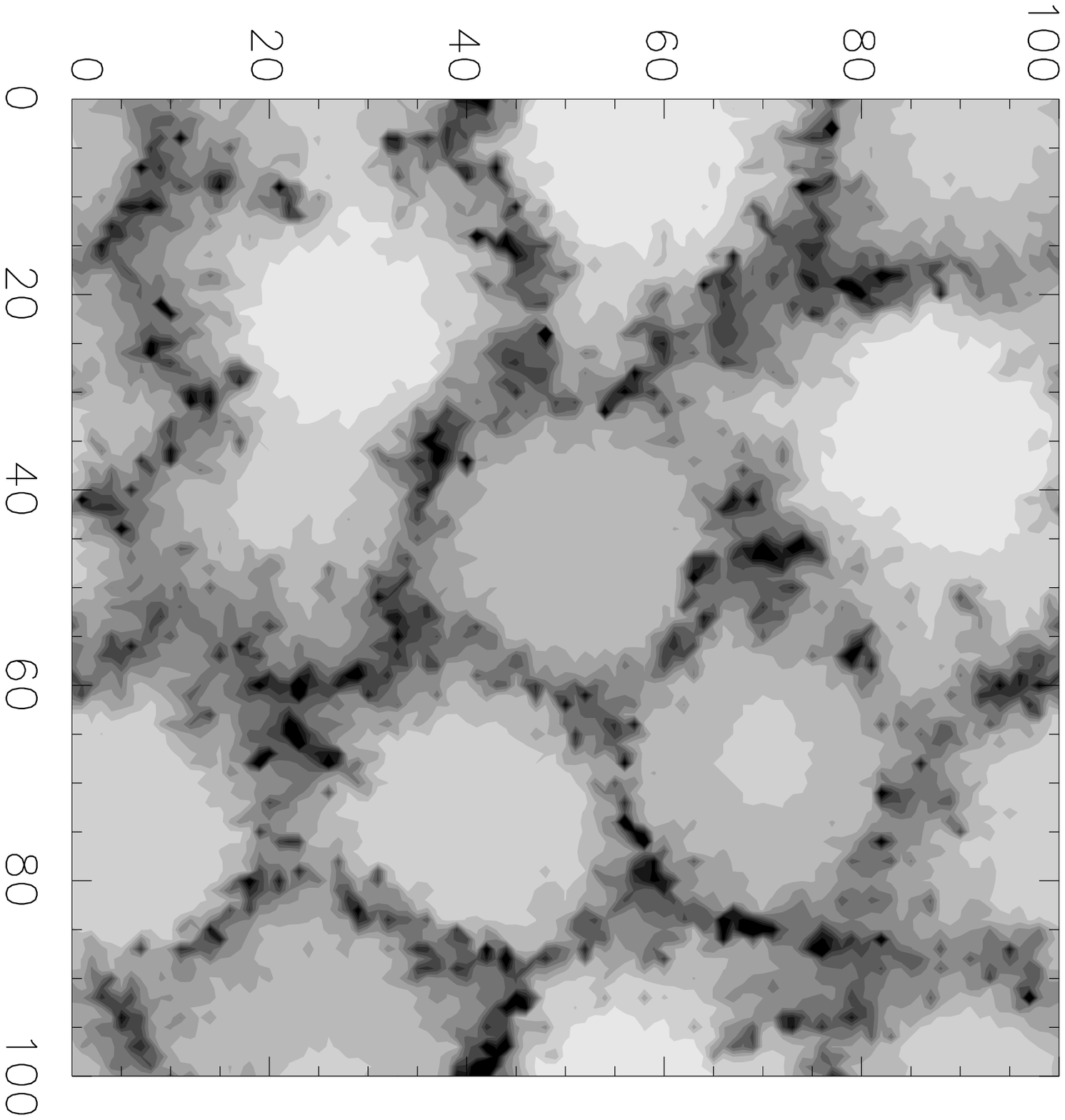}
   }
  \hss}
 }

\caption{
(a) The mound evolution in 1+1 dimensions
showing a section of $500$ middle lattice sites from a
substrate size of $L=10000$ at $10^4$ ML, $10^5$ ML, and $10^6$ ML
($P_U=1;P_L=0.5$).
(b,c) The d=2+1 growth morphologies on a $100 \times 100$ substrate
(b) $10^3$ ML, and
(c) $10^6$ ML. The darker (lighter) shades represent lower (higher)
points on the surface.
}
\end{figure}
In Fig.2 we show morphological evolutions for d=2+1 and 1+1
in order to visualize the
coarsening/steepening dynamics of the mound structures.  
We mention that in producing our final  
results we utilize a noise
reduction \cite{28} technique which accepts only a fraction of the
attempted kinetic events, and in the process produces smoother results
(reducing noise effects)
without in any way affecting the critical exponents.
The corresponding results without noise reduction are visually more
noisy with identical growth exponents.

To proceed quantitatively we now introduce the dynamic scaling ansatz
which seems to describe well all our simulated results. We have studied
the root mean square surface width or surface roughness ($W$), the
average mound size ($R$), the average mound height ($H$), and the
average mound slope ($M$) as functions of growth time. We have also
studied the various moments of dynamical height-height correlation
function, and these correlation function results (to be reported elsewhere)
are consistent with the ones obtained from our study of $W(t)$, $R(t)$,
$H(t)$, and $M(t)$. The dynamical scaling ansatz in the context of the
evolving mound morphologies can be written as power laws in growth time
(which is equivalent to power laws in the average film thickness) : $W(t)
\sim t^\beta$ ; $R(t) \sim t^n$ ; $H(t) \sim t^\kappa$ ; $M(t) \sim
t^\lambda$ ; $\xi(t) \sim t^{1/z}$, where $\xi(t)$ is the lateral correlation
length (with $z$ as the dynamical exponent) and $\beta$, $n$, $\kappa$,
$\lambda$, $z$ are various growth exponents which are not necessarily
independent. In the steady state, when 
$\xi(t) \stackrel{\displaystyle >}{\sim} L$ where $L$ is the
lateral substrate size, effective $\beta$ vanishes as the surface
roughness saturates to a value $W_s(L) \equiv W(L,t \rightarrow \infty)
\sim L^\alpha$, where $\alpha = \beta z$ is the roughness exponent. We
find in all our simulations $n \approx z^{-1}$, and thus the coarsening
exponent $n$, which describes how the individual mound sizes increase in
time, is the same as the inverse dynamical exponent in our model. We
also find $\beta = \kappa$ in all our results, which is understandable
in a mound-dominated morphology. In addition, all our results satisfy
the expected exponent identity $\beta = \kappa = n + \lambda$ because
the mound slope $M \sim H/R$. The evolving growth morphology is thus
completely defined by two independent critical exponents $\beta$ (the
growth exponent) and $n$ (the coarsening exponent), which is similar to
the standard (i.e. without any diffusion bias) dynamic scaling situation
where $\beta$ and $z$ ($= n^{-1}$ in the presence of diffusion bias)
completely define the scaling properties. We note also that our {\it
negative} bias results (not shown here for lack of space) 
are completely consistent with the expected
\cite{1,22} Edwards - Wilkinson universality class 
\cite{1,21,22} with our numerical
findings being $\beta = 0.25$ and $z=2$ in d=1+1, and $\beta = 0$
(i.e. $W \sim \ln t$) and $z=2$ in d=2+1.
This is expected because our negative bias model explicitly introduces 
an inclination dependent
downhill surface current \cite{29} in the problem. 

In Fig.3 we show our representative scaling results for nonequilibrium
growth under surface diffusion bias conditions. It is clear that we
consistently find $\beta \simeq 0.5$ in both d=1+1 and 2+1 for 
growth under a surface diffusion bias.
\begin{figure}

 \vbox to 5.2cm {\vss\hbox to 6cm
 {\hss\
   {\includegraphics{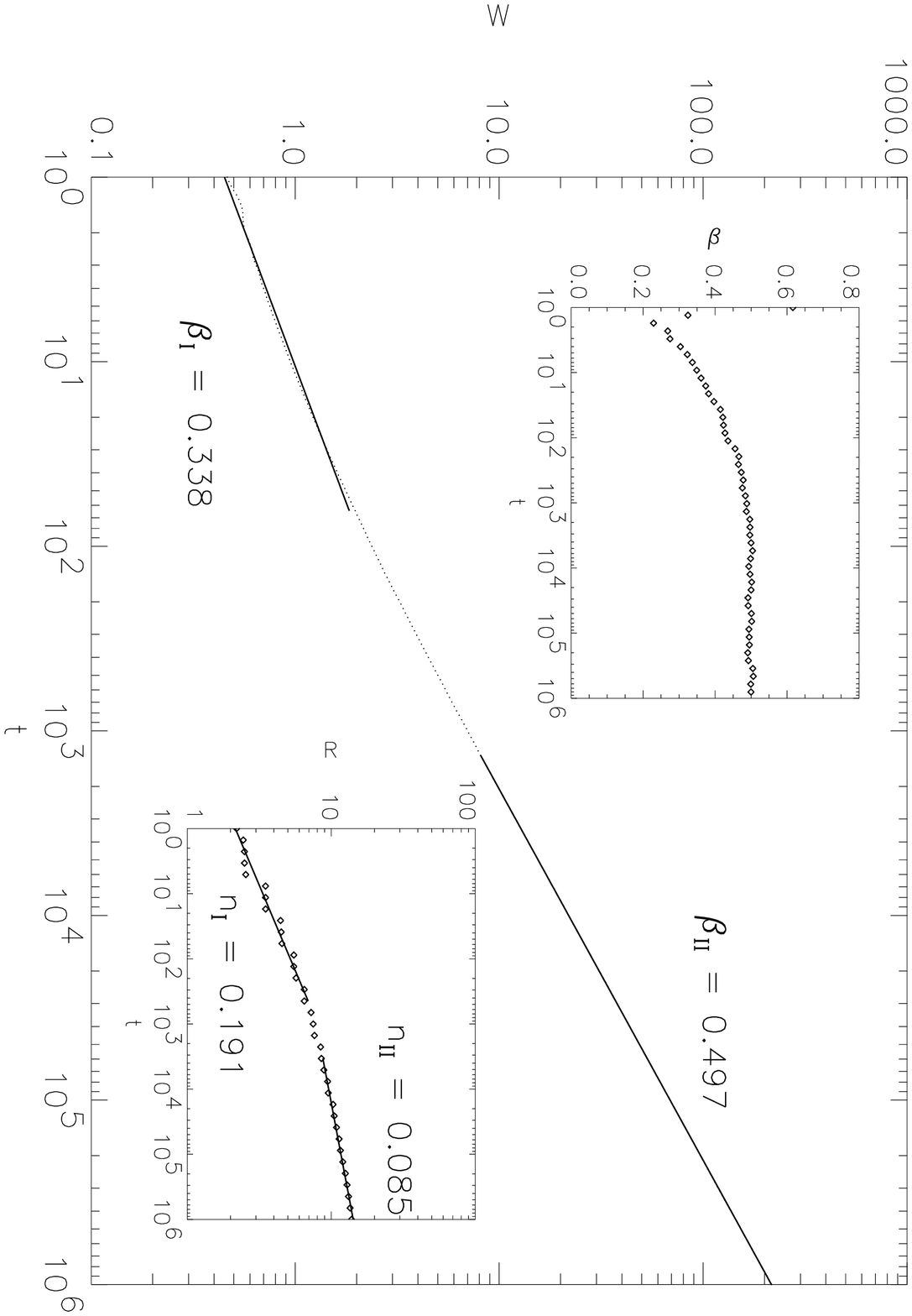}
   }
  \hss}
 }

 \vbox to 5.2cm {\vss\hbox to 6cm
 {\hss\
   {\includegraphics{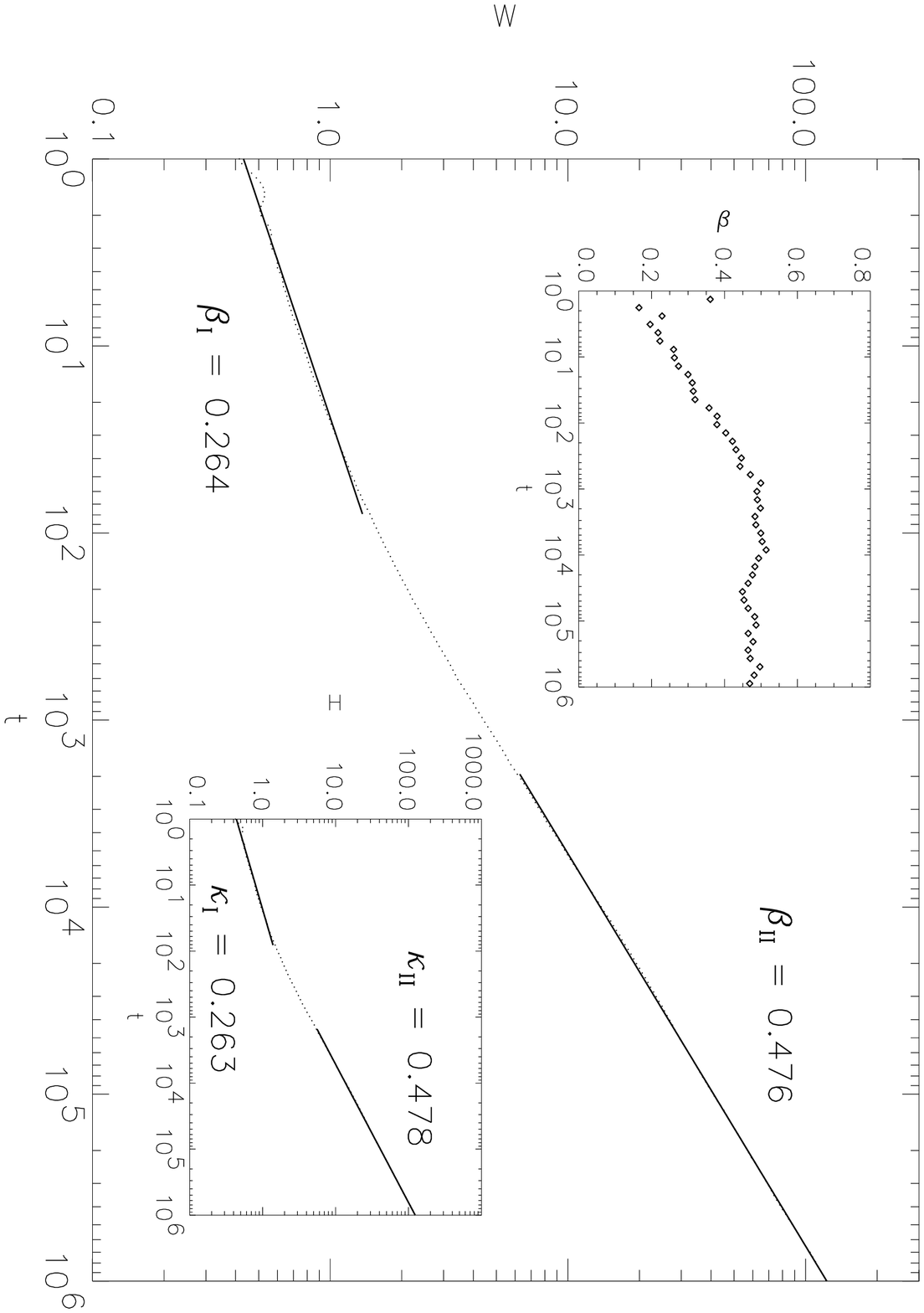}
   }
  \hss}
 }

 \vbox to 5.2cm {\vss\hbox to 6cm
 {\hss\
   {\includegraphics{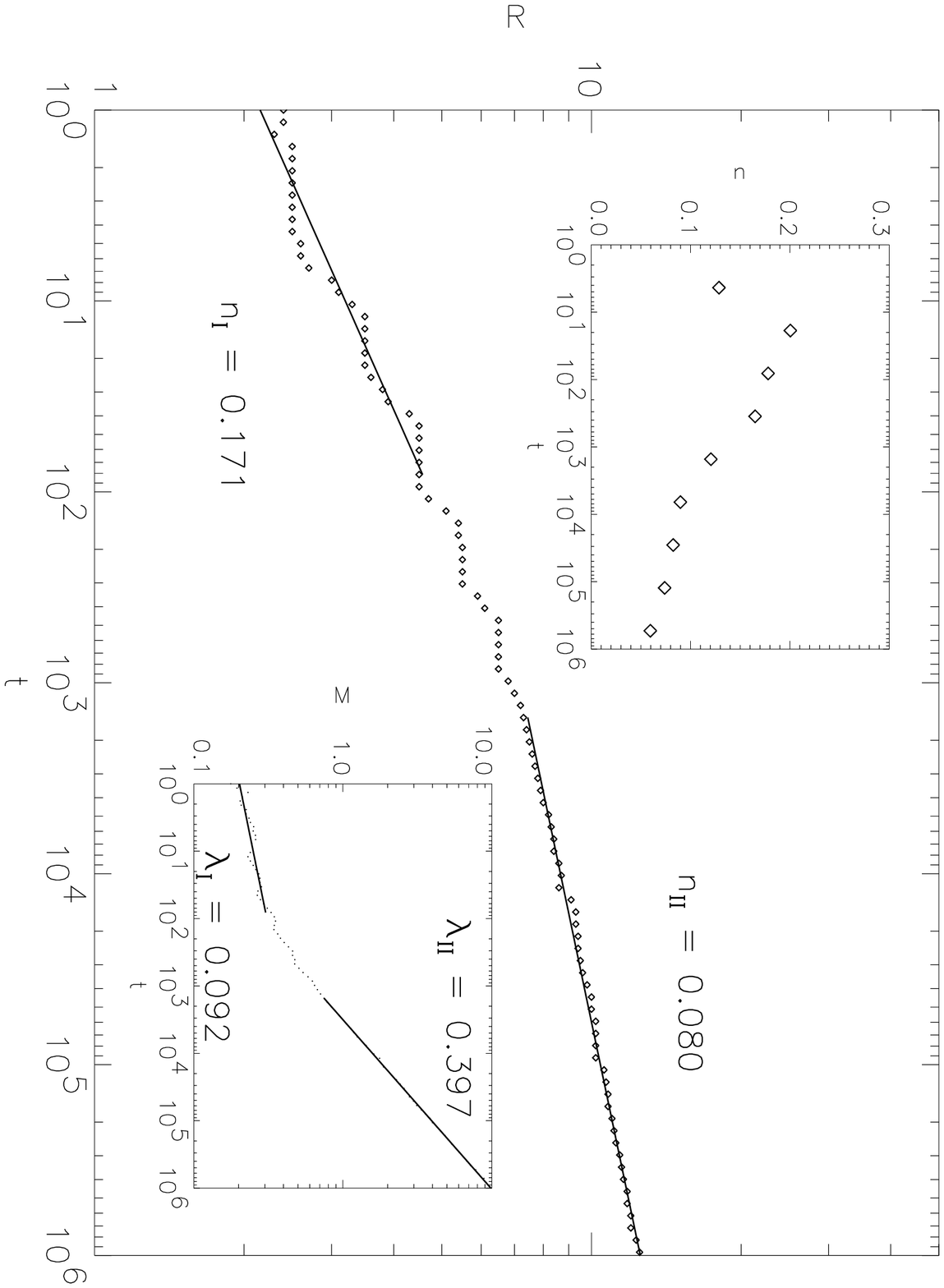}
   }
  \hss}
 }

\caption{
(a) The surface roughness $W$ in 1+1 dimensions
as a function of deposition time $t$
in the $L=10000$ system. Left inset: the growth exponent $\beta$
calculated from the local derivative of $\log_{10}W$ with respect to
$\log_{10}t$. Right inset: Average mound size as a function of
deposition time.
(b) The surface roughness in the $100 \times 100$ system. Left inset:
the local growth exponent $\beta$. Right inset: Average mound height vs
time.
(c) The average mound size in the $100 \times 100$ system. Left inset:
the local coarsening exponent $n$ calculated from the local derivative
of $\log_{10}R$ with respect to $\log_{10}t$. Right inset: the average
mound slope vs time.
}
\end{figure}
This $\beta \simeq 0.5$ is, however, very different from the usual
Poisson growth under pure random deposition with no relaxation
where there are no lateral growth correlations.
Our calculated asymptotic coarsening exponent $n$ in both d=1+1 and
2+1 is essentially zero ($< 0.1$) at long times.
In all our results we find the effective coarsening exponent showing a
crossover from $n \approx 0.2$ at early times to a rather small value
($<0.1$) at long times -- we believe the asymptotic $n$ to be zero in
our model.

In comparing with the existing continuum growth equation results (of
which there are quite a few) we find that none can quantitatively
explain all our findings. 
Golubovic \cite{18} predicts $\beta = 0.5$, which is consistent with our
simulations, but his finding of $n = \lambda = 1/4$ in both d=1+1 and
2+1 is inconsistent with our results 
($n < 0.1$, $\lambda \simeq 0.4-0.5$)
except in the limited transient regime. The analytic results of
Rost and Krug \cite{19} also cannot explain our results, because they
predict, in agreement with Golubovic, that if $\beta = 1/2$, then $n =
\lambda = 1/4$, which is in disagreement with our findings.
We also find our $\beta$ to be essentially $0.5$ independent of the
actual value of $P_U - P_L$, which disagrees with ref. 19.
Our model obviously has no slope selection ($\lambda \equiv 0$), and
therefore theories predicting slope selections do not apply.

Before concluding we point out some reasons for why we believe our
nonequilibrium limited mobility
growth model to be a good zeroth order description for
kinetic growth under a surface diffusion bias. The main reason for this
is the success of the corresponding minimal growth model, introduced in
ref. 23, in providing a good zeroth order description of molecular beam
epitaxial growth in the absence of any surface diffusion bias. The d=2+1
critical exponents \cite{27} in the unbiased model \cite{23} are
$\beta = 0.25 - 0.2$ and $\alpha \simeq 0.6-0.7$, which are in 
quantitative agreement with a number of experimental measurements
\cite{21,22} where surface diffusion bias is thought to be 
dynamically unimportant.
An equally significant theoretical reason is that the
corresponding unbiased growth model 
\cite{23,24,25,26,27} is the {\it only
existing} nonequilibrium growth model which is known \cite{22,27} {\it
not} to have the linear Edwards - Wilkinson `$\nabla ^2 h$' term
\cite{1} in its coarse-grained long wavelength continuum description
(an equivalent statement \cite{29} is that this is the only 
limited mobility model which
has a vanishing surface current on a tilted substrate) -- in fact, our
negative bias model introduces precisely this `$\nabla ^2 h$' term by
producing a downhill current on a tilted substrate. 
(The other limited mobility nonequilibrium growth models \cite{30,31}
introduced in the literature are known to belong to the Edwards -
Wilkinson universality class, and are therefore unsuitable for diffusion
bias studies \cite{32}.)
Therefore, the
minimally biased version of this model which we study in this paper
should be the appropriate zeroth order description for nonequilibrium
growth under surface diffusion bias conditions. Since an approximate
continuum description for the unbiased model \cite{23}
has recently been developed
\cite{26}, one could use that as the starting point to construct a
continuum growth model for the biased growth situation. 
Such a continuum description is, however, extremely complex \cite{26} as
it requires the existence of an infinite number of terms in the growth
equation. It remains unclear that a meaningful continuum description for
nonequilibrium growth under a surface diffusion bias is indeed possible
even for our simple limited mobility growth model.

We conclude by mentioning that any comparison to experiments, as has been
done in several recent theoretical publications on this topic, is
premature at this stage because the real Ehrlich - Schwoebel barrier in
experimental systems \cite{33} is expected \cite{34} to be quite
complicated, and simple growth models used by us and others
\cite{4,5,6,7,8,9,10,11,12,13,14,15,16,17,18,19,20} most likely do not
apply. What we have established in this paper is that even a very simple
limited mobility nonequilibrium growth model leads to extremely complex
dynamical interface morphologies under surface diffusion bias
conditions. The fact that even a deceptively simple limited mobility
growth model such as the one studied in this paper seems to defy a
theoretical continuum description is a strong indication that our
understanding of nonequilibrium growth under a surface diffusion bias is
far from complete. 

This work is supported by the US-ONR and the NSF-MRSEC.

\end{document}